\documentclass[usenatbib,conference]{basi}
\usepackage{graphicx,epsfig,amsmath,amssymb,amsbsy,natbib}
\usepackage[T1]{fontenc}
\usepackage[british]{babel}
\usepackage[varg]{txfonts}
\usepackage{rotating}
\usepackage{dcolumn}
\begin{document}
\title[MHD of accretion mounds]{MHD instabilities in accretion mounds on neutron star binaries}
\author[Mukherjee, Bhattacharya and Mignone]%
       {Dipanjan Mukherjee$^1$\thanks{email: \texttt{dipanjan@iucaa.ernet.in}},
       Dipankar Bhattacharya$^{1}$ and Andrea Mignone$^2$\\
       $^1$Inter University Center for Astronomy and Astrophysics, post bag 4, Ganeshkhind, Pune, India.\\
       $^2$Dipartimento di Fisica Generale, Universita di Torino, via Pietro Giuria 1, 10125 Torino, Italy.}

\pubyear{2013}
\volume{00}
\pagerange{\pageref{firstpage}--\pageref{lastpage}}

\date{Received --- ; accepted ---}

\maketitle

\label{firstpage}

\begin{abstract}
We have numerically solved the Grad-Shafranov equation for axisymmetric static MHD equilibria of matter confined at the polar cap of neutron stars. From the equilibrium solutions we explore the stability of the accretion mounds using the PLUTO MHD code. We find that pressure driven modes disrupt the equilibria beyond a threshold mound mass, forming dynamic structures, as matter spreads over the neutron star surface. Our results show that local variation of magnetic field will significantly affect the shape and nature of the cyclotron features observed in the spectra of High Mass X-ray Binaries. \\[6pt]
\end{abstract}


\section{Introduction}\label{s:intro}
Neutron stars in binary systems accrete matter from the companion star, channelling the matter towards the poles. The accreted matter is confined in a mound by the polar magnetic field. Distortions in the local magnetic field due to the pressure of the accreted matter can significantly affect the cyclotron resonance scattering features (CRSF) formed there. In the long term such field distortions may contribute to field burial through diamagnetic screening \citep{romani90,melatos01,melatos04}, but the extent of this may be limited by MHD instabilities (e.g. Litwin et~al. 2001). 

In this presentation, we first present the solutions of the magnetostatic equations describing the accretion mound. We show that accreted matter distort the field lines from the unloaded dipolar structure, even at heights much larger than that of the mound itself. Next, we perturb the static solutions to study the stability of the system and the growth of MHD modes. We find that for mounds above a threshold mass, MHD instabilities disrupt the equilibria. We discuss the implications of the local field distortions on the cyclotron resonance scattering features (CRSF) emitted from such systems and the effect of the instabilities on the long term evolution of the system.

\section{Magnetostatic solutions of accretion mounds}
We consider an accretion mound of polar cap radius $R_p\sim 1$~km on a neutron star of radius $\sim 10$~km, mass $\sim 1.4\; M_\odot$ and polar surface field strength $\sim 10^{12}$~G, typical of mounds on HMXB systems. We consider Newtonian gravity of constant acceleration $\boldsymbol{g}=-g \boldsymbol{\hat{z}}$. We work in a cylindrical coordinate system ($r,\theta,z$) with the origin at the magnetic pole and assume axisymmetry around the $z$ axis. By introducing the flux function $\psi(r,z)$ describing the poloidal flux through a circle of radius $r$ at a given height, one can recast the static Euler equation into the Grad-Shafranov (hereafter GS) equation \citep{dipanjan12}:
\begin{equation}
\frac{\Delta ^2 \psi}{4\pi r^2} = -\rho g \frac{dZ_0}{d\psi}
\end{equation}
Previously, the approximate equation of state for a non-relativistic degenerate Fermi gas (with $p \propto \rho ^{5/3}$) was used to solve the GS equation, which is insufficient to describe the plasma for large densities near the base ($\geq 10^6 \mbox{ g cm}^{-3}$). In our current work, we have used an equation of state: $p= (8 \pi/15)m_ec^2\left(\frac{m_ec}{h}\right)^3 x_F^5/\left((1+16/25x_F^2)^{1/2}\right)$, which closely approximates the $T=0$ K Fermi plasma (with errors less than $\sim 1.5\%$; \citet{pacz83}). Here $X_F=\frac{1}{m_e c} \left(\frac{3 h^3}{8 \pi \mu _e m_p}\right)^{1/3} \rho ^{1/3}$ is the Fermi momentum. The density can be derived from the expression for Fermi momentum obtained after separation of variables:
\begin{equation}
x_F = \frac{5}{4}\left( \frac{\xi ^2 - 8/3 + \xi \sqrt{16/9 + \xi ^2}}{32/9} \right) ^{1/2} \; ; \; \xi=\frac{16}{15} \frac{\mu _e m_p}{m_e c^2} \left(Z_0(\psi) - z\right) + 1 \end{equation}
The shape of the mound is specified by a mound height function $Z_0(\psi)$ marking the top of the mound as a function of flux.
\begin{figure}
\centerline{
\includegraphics[height=6.5cm,width=5cm,keepaspectratio]{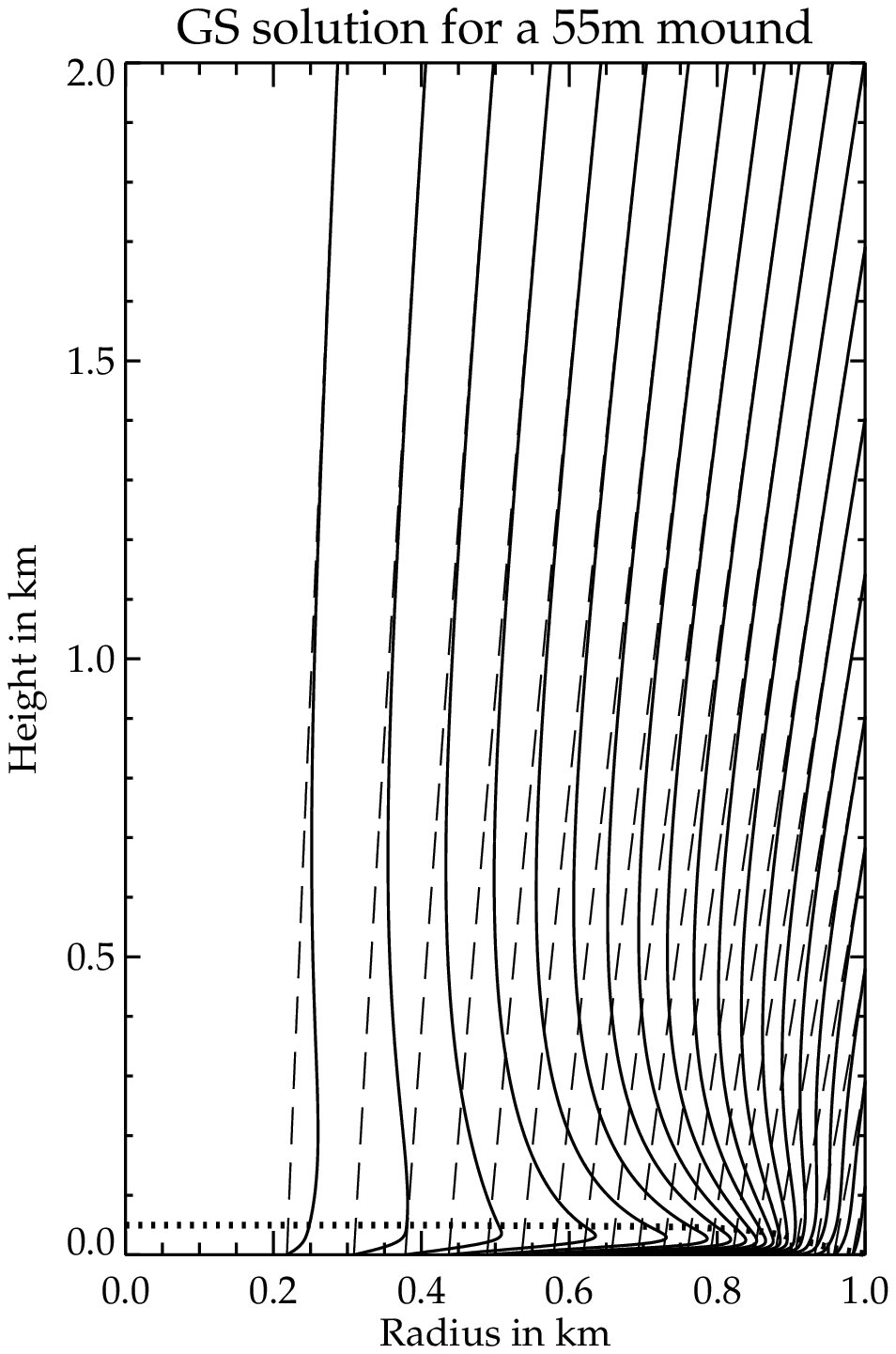}
\includegraphics[height=6.5cm,width=5cm,keepaspectratio]{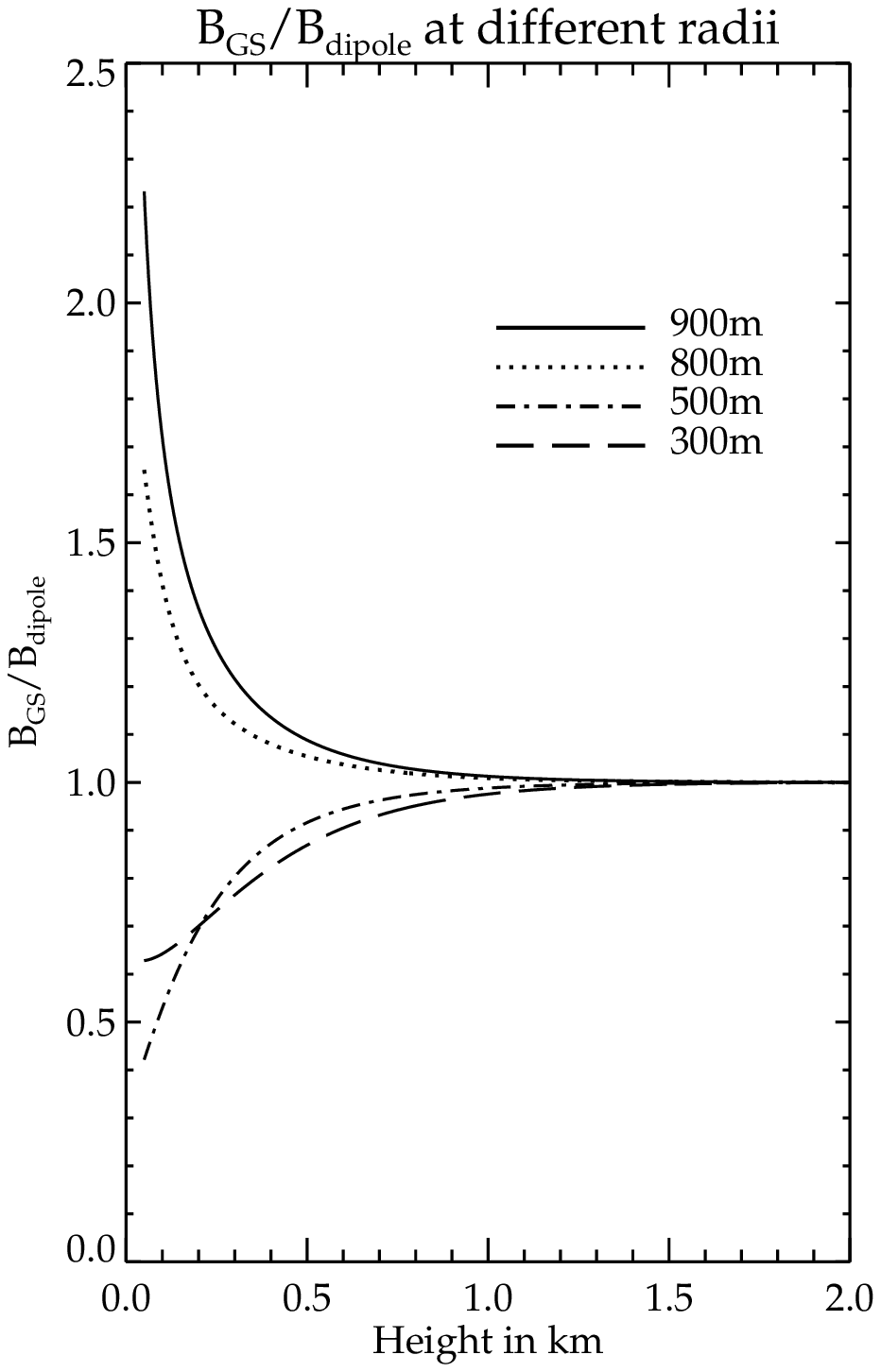}
}
\caption{\small Left: The field lines inside the column for a GS solution of a 55m mound (solid lines), compared with undistorted dipolar field (dashed). The dotted line represents the top of the mound. Substantial deviation from dipolar fields extends far above the mound surface, up to a height $\sim 1$km. Right: Ratio of strength of the local field to that of an undistorted dipole, as a function of height, at different radial distances from the magnetic axis. For $r \geq 700$m (where pressure gradients are highest), field lines are pushed outside by confined matter causing enhancement of field strength, and a decrease in the inner parts. Even at a height $\sim 500$m, field strength differs by more than 10\% of dipole value.}\label{GSsol}
\end{figure}
Solutions thus obtained show large deviation from dipolar field configuration, even at heights several hundred metres above the mound (see Fig.~\ref{GSsol}). CRSF emitted from such columns  will have complex shapes and features \citep{dipanjan12}. Some sources like V0332+53 show broader CRSF with decrease in luminosity \citep{tsygankov10} which can be interpreted as the characteristic emission region coming closer to the mound where field distortion is larger. Current and future X-ray missions with improved spectral sensitivity like NuSTAR, ASTROSAT, ASTRO-H, LOFT etc will be crucial to probe the conditions inside such accretion columns.

\section{MHD instabilities in accretion mounds}
\begin{figure}
\centering
\includegraphics[height=5cm,width=5cm,keepaspectratio]{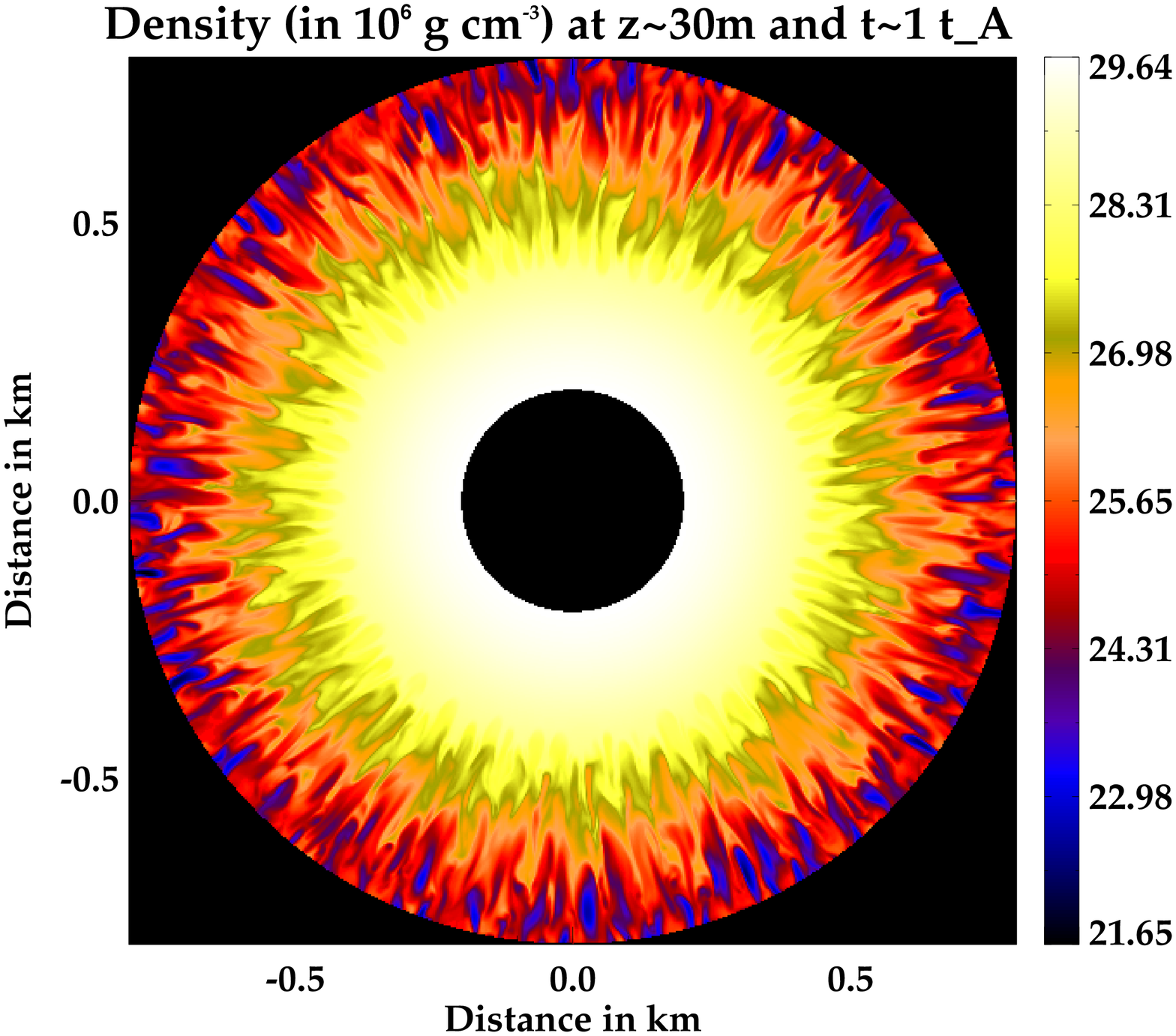}
\includegraphics[height=5cm,width=5cm,keepaspectratio]{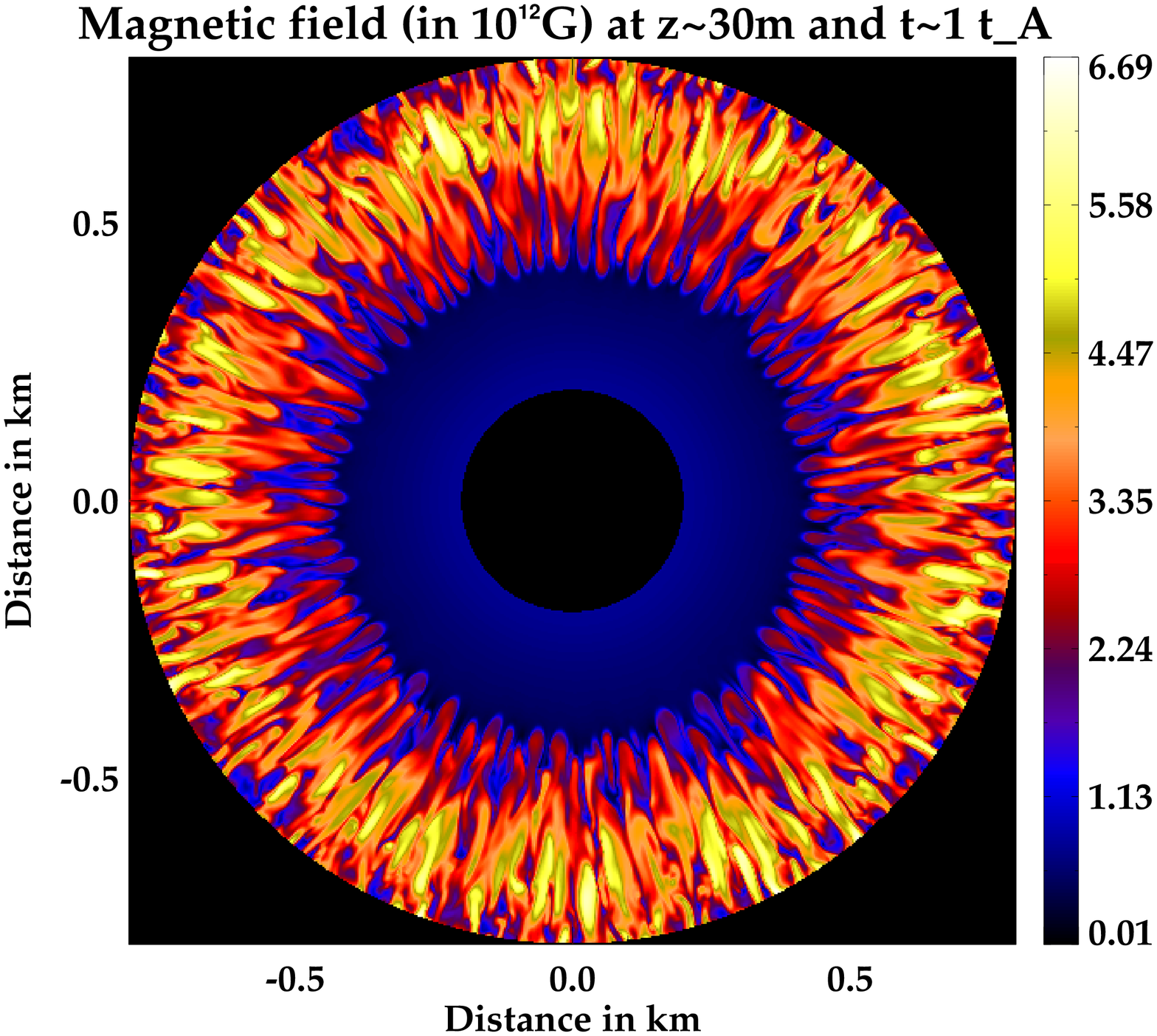}
\caption{\small Left: Cross section of a 70m mound at a height of $\sim 30$m and $t\sim 1t_A$ showing the density. $t_A$ is the average local Alfv\'en time $\sim 2.8 \times 10^{-3}$s. Right: The magnetic field magnitude at the same height and time. The finger like channels due to the MHD instabilities are clearly seen at the outer radial edges. See the online journal for a colour version of the figure.}\label{PLUTO}
\end{figure}
To investigate the presence of MHD instabilities, we perturb the GS equilibrium solution and follow the dynamics with the PLUTO MHD code (Mignone et~al. 2007). Mukherjee et.~al 2013 have shown the presence of gravity driven modes through 2D axisymmetric simulations. Here we report the results from 3D non-axisymmetric simulations of the mounds.  We use GS solutions with $p \propto \rho ^{5/3}$ equation of state, for easier numerical implementation of the MHD equations. We have performed these simulations for mounds of different shapes and masses to study the effect on the growth rates of the MHD instabilities. Details of the numerical simulations will be presented in a forthcoming paper (Mukherjee, Bhattacharya and Mignone, submitted to MNRAS).

Mounds of larger mass ($\sim 10^{-12} M_\odot$) with larger field curvatures are highly prone to pressure driven instabilities. Finger like channels appear at the radial edges in a few Alfv\'en times as matter passes through regions of low magnetic fields (see Fig.~\ref{PLUTO}). The instabilities develop quickly over time scales of milliseconds. However for mounds of smaller mass with less field curvature (e.g. 50m mound with mass $\sim 6.8 \times 10^{-13} M_\odot$), the growth time scales are ten times slower. Eventually we reach a threshold mound size ($\sim 45$m mound of mass $\sim 5\times 10^{-13} M_\odot$) which is stable to perturbations. The maximum plasma $\beta$ (ratio of plasma pressure to magnetic pressure) for a 45m mound is $\sim 293$ which is close to the threshold $\beta \sim 260$ predicted by Litwin et~al. (2001) for instability. 

Previous solutions of \citet{melatos01} and \citet{melatos04} predict large accretion mounds on the neutron stars formed due to continued accretion over long time scales. Such large mounds drag the field lines to form local screening currents. However MHD instabilities as presented here, will severely limit such screening currents from being formed. Presence of instabilities in much smaller mound sizes than previous estimates indicates that such MHD processes will play an important role in determining the long term evolution of the field and the spread of the accreted matter.

\nocite{litwin01}
\nocite{tsygankov}
\nocite{andrea07}
\nocite{dipanjan13a}
\def\apj{ApJ}%
\def\mnras{MNRAS}%
\def\aap{A\&A}%
\def\apjl{ApJ}
\def\physrep{PhR}
\def\apjs{ApJS}
\def\pasa{PASA}
\def\pasj{PASJ}
\def\nat{Nature}
\def\memsai{MmSAI}

\bibliographystyle{mn2e}
\bibliography{dipanjanbib}

\end{document}